*Review*

# A Review of Landcover Classification with Very-High Resolution Remotely Sensed Optical Images—Analysis Unit, Model Scalability and Transferability

**Rongjun Qin** [1,2,3,4] **and Tao Liu** [5,6,*]

[1] Geospatial Data Analytics Lab, The Ohio State University, 218B Bolz Hall, 2036 Neil Avenue, Columbus, OH 43210, USA; qin.324@osu.edu
[2] Department of Civil, Environmental and Geodetic Engineering, The Ohio State University, 2070 Neil Avenue, Columbus, OH 43210, USA
[3] Department of Electrical and Computer Engineering, The Ohio State University, 205 Dreese Labs, 2015 Neil Avenue, Columbus, OH 43210, USA
[4] Translational Data Analytics Institute, The Ohio State University, Pomerene Hall, 1760 Neil Ave,
Columbus, OH 43210, USA
[5] College of Forest Resources and Environmental Science, Michigan Technological University, 1400 Townsend Drive, Houghton, MI 49931, USA
[6] Ecosystem Science Center, Michigan Technological University, 1400 Townsend Drive,
Houghton, MI 49931, USA
[*] Corresponding: taoliu@mtu.edu

**Abstract:** As an important application in remote sensing, landcover classification remains one of the most challenging tasks in very-high-resolution (VHR) image analysis. As the rapidly increasing number of Deep Learning (DL) based landcover methods and training strategies are claimed to be the state-of-the-art, the already fragmented technical landscape of landcover mapping methods has been further complicated. Although there exists a plethora of literature review work attempting to guide researchers in making an informed choice of landcover mapping methods, the articles either focus on the review of applications in a specific area or revolve around general deep learning models, which lack a systematic view of the ever advancing landcover mapping methods. In addition, issues related to training samples and model transferability have become more critical than ever in an era dominated by data-driven approaches, but these issues were addressed to a lesser extent in previous review articles regarding remote sensing classification. Therefore, in this paper, we present a systematic overview of existing methods by starting from learning methods and varying basic analysis units for landcover mapping tasks, to challenges and solutions on three aspects of scalability and transferability with a remote sensing classification focus including (1) sparsity and imbalance of data; (2) domain gaps across different geographical regions; and (3) multi-source and multi-view fusion. We discuss in detail each of these categorical methods and draw concluding remarks in these developments and recommend potential directions for the continued endeavor.

**Keywords:** very-high resolution; VHR, landcover classification; semantic segmentation; analysis unit; deep learning; transfer learning; data fusion; remote sensing

## I. Introduction

Landcover mapping using remote sensing (RS) images has presented a consistent requirement for decades since the collection of the very first RS image. It greatly facilitates automated analysis of urban, suburban, and natural environments for applications such as urban expansion monitoring, change detection, crop prediction, forestation/deforestation, surveillance, anthropogenic activities, mining, etc. Generally, it is considered as a highly disparate problem and often appears to be application- and even location-dependent, as the learning systems need to accommodate the varying reference/training data with quality and availability, the complexity of landcover classes, and the multi-source/multi-modal datasets. As of now, the accurate production of coarse resolution landcover classification maps at the global level (e.g., 30-m resolution) still follows a semi-automated approach and is often labor-intensive [1]. However, when it comes to high-resolution or very-high-resolution (VHR) images (resolution at the sub-meter level), there exist even greater challenges in classification tasks, due to the desired high-resolution output and level of uncertainty in predictions. Presently, with the collection of imageries from spaceborne sensors/platforms such as WorldView constellations, IKONONS (decommissioned), GeoEye, Pleiades, Planet, etc., the volume of available VHR images has increased to an unprecedented level, and there exist a large body of approaches developed to address the classification of VHR data, from simple statistical learning based spectrum classification, spatial–spectral feature extraction, towards the recently popularized deep learning (DL) methods. Additionally, the available training data in varying multi-source forms (e.g., multi-view, multi-temporal, LiDAR data, or Synthetic Aperture Radar data), are becoming decisive when considering the scalability of landcover classifications.

Since the introduction of DL to the RS community, it has been widely adopted to solve a variety of RS tasks in classification, segmentation and object detection, and a few relevant review articles were published relevant to DL-based RS applications. For example, [2] reviewed the DL models for road extraction during the time period from 2010 to 2019; [3] discussed data sources, data preparation, training details and performance comparison for DL semantic segmentation models for satellite images in urban environments; [4],[5] reviewed DL applications in hyperspectral and multispectral images; [6],[7] reviewed DL approaches to process 3D point cloud or RS data; [8] reviewed various applications based on DL models including detection and segmentation of individual plants and vegetation classes; [9] broadly reviewed applications of DL models on various RS tasks including image preprocessing, change detection and classification; [10] discussed various DL models used for wetland mapping; [11],[12] reviewed 429 studies to investigate the impact of DL for earth observation applications through image segmentation and object detection; [13] reviewed 416 papers in a meta-analysis and discussed the distribution of data sensors, DL models and application types in those studies; [14,15] reviewed the deep learning applications for scene classification on aspects of the challenges, methods, benchmarks and opportunities.

While these literature surveys mainly focused on the diverse applications and the achievable accuracies of different models, few summarized works that inherently lead to scalable solutions to address the problem of training data sparsity of domain gaps. Moreover, existing reviews on RS classification, primarily introduced DL as a general approach or scene classification method, thus there remains a lack of a systematic taxonomy of DL methods specifically focusing on landcover mapping. In this paper, we aim to provide a comprehensive review of landcover classification methods from the perspective of the ever-growing data and scalability challenges. To begin, we first draw the connections between the DL-based approaches and traditional classifiers in terms of the analysis unit, to form an easily understandable taxonomy of DL approaches. Following these basics, we then outline its existing scalability challenges and summarize available solutions addressing them, and present our views on these potential solutions.

*1.1. Scope and Organization of This Paper*

In this paper, we aim to introduce existing works addressing issues (e.g., low quantity and quality of training samples, domain adaption, multimodal data fusion, etc.) related to landcover mapping using VHR images and outlook potential research directions. While our review focuses on works with applications to VHR RS data (data with Ground Sampling Distance of 2 m or less), we might introduce representative works on lower-resolution data when the relevant methods are of value to support VHR data for this review. Furthermore, although this review will largely involve the new opportunities and methods brought by DL, we maintain an equivalent emphasis on the relevant shallow classifiers as well, especially when addressing framework-level approaches (semi-/weak- supervision) that are model agnostic. Given that there is a large body of literature related to general machine learning models and methodologies, this review only addresses approaches related to or adapted to RS image classifications.

The rest of this review is organized as follows: In Sections 1.2 and 1.3, we briefly review the existing issues of landcover mapping using VHR and the related efforts to address those issues; In Section 2, we provide a concise illustration of landcover classification paradigms using VHR from the perspective of the analysis unit to engage necessary contents and basics; In Section 3, we elaborate on existing approaches addressing the data challenges (briefly mentioned in 1.2 and 1.3) for RS landcover classification. In Section 4, we conclude this review by discussing our findings and providing our outlooks on potential approaches to move existing practices forward.

*1.2. Existing Challenges in the Landcover Classification with VHR Images*

As the number of available VHR images, annotated data, and complexity of learning models continue to grow, there have been many landcover classification studies focusing on a diverse set of applications. Among these studies, the major challenges presented in RS image landcover classification for VHR data are mainly centered on three aspects: (1) The intra-class variability and inter-class similarity affecting classification accuracy; (2) imbalance, inconsistency and lack of quality training data for training high-accuracy classifiers; (3) large domain gaps across different scenes and geographical regions when scaling up the well-trained classifiers from a particular dataset.

1.2.1. Intra-Class Variability and Inter-Class Similarity for VHR Data

The VHR data at a resolution of a meter or less on the ground, have brought the benefits of obtaining greater details for earth observation. Along with the increased resolution, it has introduced greater intra-class variability and inter-class similarities: if using the spectral information alone, a pixel may be easily identified as belonging to multiple land-cover classes; equivalently, different classes may contain pixels with similar spectral signatures [16]. Although solutions for this particular challenge such as object-based methods or spatial–spectral features [17,18] were intensively investigated, the achievable improvements failed to keep up with the increased resolution and volume of data. As a result, it may become even more problematic when advanced (and more complex) models with increased annotated datasets are used. For example, deep-learning (DL) models bring in drastically improved accuracy for specific and well-defined tasks, as they tend to fit the varying signals of the same class and at the same time to discern the tiny difference of classes with similar spectrum information (e.g., roof vs. ground). However, as is well known, this is at the expense of generalization, which may drastically decrease the performance when applying a trained DL model to different datasets with drastically different data distributions [19,20].

1.2.2. Imbalance, Inconsistency, and Lack of Quality Training Data

The increasing volume of VHR data and complexity of models naturally demand more training data, while the traditional manual labeling approaches primarily used in the era of processing coarse resolution data (such as MODIS, Landsat, and Sentinel) [21-23] or VHR but within small AOI, are sub-optimal and no longer feasible as the models are transitioned to the more data-demanding DL models. To overcome this issue, researchers sought to collect training samples from multiple sources, including crowdsourcing services (e.g., Amazon Turk) [24], and public datasets (e.g., OpenStreetMap) [25]. These additional datasets, on one hand, possess a great effort-reducing value for training high-accuracy classifiers, while on the other hand, introduce additional challenges that may need solutions for common training data issues detailed in the following.

*Imbalanced training samples*: Imbalanced training samples are often associated with the scene, as the number of training samples per class might not be necessarily the same and can be scene dependent. This was somehow inherently handled in traditional manual labeling approaches, as samples were purposefully drawn and post-resampled for shallow classifiers. For DL-based models, often all available training data are fed into a network regardless of their balance; therefore, appropriate strategies both in data augmentation, training and testing are required to accommodate the imbalanced training data problems.

*Inconsistency of training samples*: There has been a recent boost of crowdsourcing or public datasets for researchers in the RS community to perform semantic segmentation [26-28]. However, these crowdsourcing datasets or public benchmark datasets may come with inconsistent class definitions and the level of details. For example, some datasets consider buildings as part of the urban class consisting of all man-made objects on the ground, while some others consider a more detailed classification that separates buildings from other man-made structures (an example is shown in Figure 1a); some datasets define the ground class as an inclusive class that contains low-vegetations, grass, and barren land, and some others separate the ground into range-land with low-vegetation, and barrens and agricultural field as separate classes (an example is shown in Figure 1b). Therefore, the first challenge to harness the use of such data is on how to adapt or refine their labels to fit specific needs and details on the classification tasks.

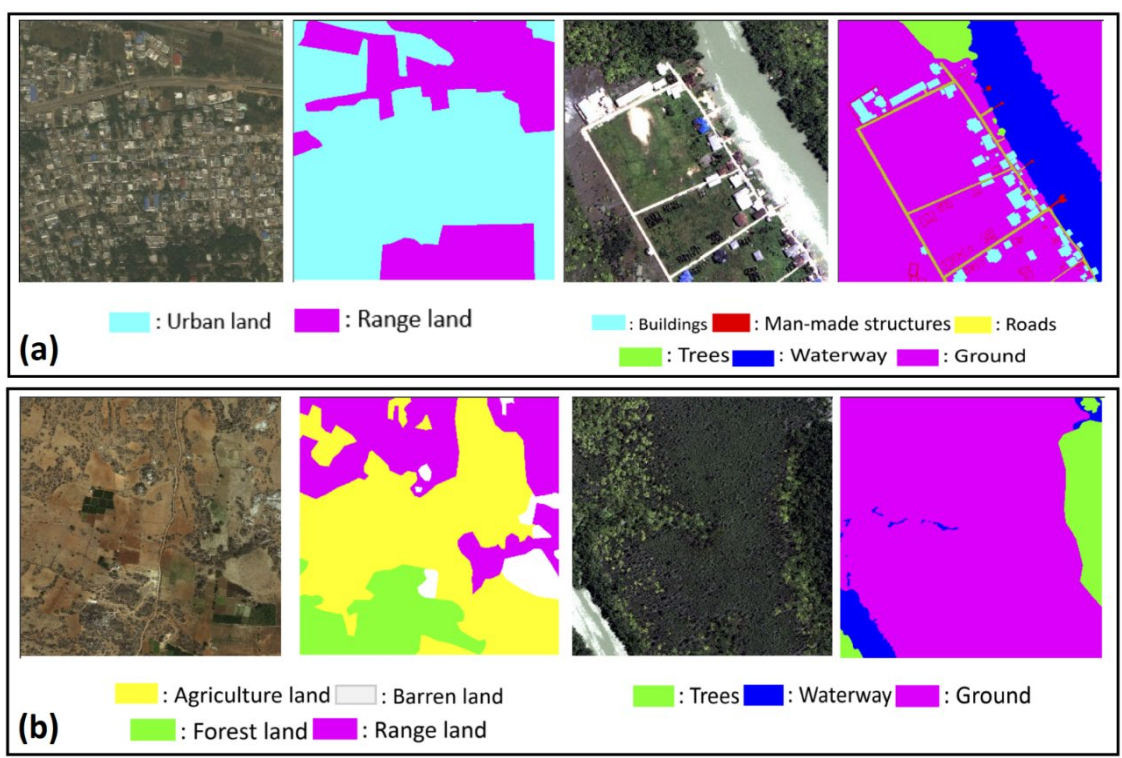

**Figure 1.** Sample image and label patches from publicly available benchmarks showing inconsistencies of the class definition and level of details, leading to challenges of using them directly as training sets for various RS classification tasks. In (**a**), different benchmark dataset shows different levels of details (Left: A sample patch from DeepGlobe dataset [27]; right: a sample patch from the DSTL dataset [29]) and (**b**) different benchmark dataset shows different class definitions (Left: A sample patch from DeepGlobe dataset [27] with no specific definition of the ground; right: A sample patch from DSTL dataset [29] defines the ground class but include many low vegetations and, inaccurately, some high vegetations).

*Lack of quality training data*: As indicated in [30], the accuracies of the most existing machine learning models in RS are underestimated, often as a result of being polluted by imperfect and low-quality training data. This presents as a common issue although active efforts are being taken e.g., to feed the community more data as samples, thus the low-quality data assumed for learning algorithms can present another challenge.

1.2.3. Model and Scene Transferability

Transferability is often a desired feature of trained models, i.e., even the test data are captured from different sensors or different geographical locations with different land patterns compared with training data, the model will yield satisfactory performance as though it were applied to the source dataset (where the training data were collected). However, the domain gaps in RS images are often underestimated [30]. Many computer vision applications claimed good transferability at the task level [31], for example, semantic segmentation or depth estimation of outdoor crowdsourcing images [32,33] are regarded to generalize well, which are somehow inherently determined by the structure of the scene from a ground-view image: the lower part is ground, left and right sides are the façade of buildings or road extended to skylines and the upper part of the images are mostly sky. Whereas in RS images, the content of different parts of the images may come with a large variation and thus are completely unstructured, in addition to which the atmospheric effects create even larger variations on the object appearances, let alone the drastic change of land patterns across the different geographical region (e.g., urban vs. suburban, tropical area vs. frigid area, etc.). It is well-noted that every single RS image could be a domain [34,35]. Therefore, to scale up classification capabilities, transferability issues remain one of the main challenges to face.

*1.3. Efforts of Harnessing Novel Machine Learning Applications and Multi-Source Data under the RS Contexts*

The above mentioned challenges represent the major barriers in modern VHR RS image classification. In addition to enhancing model performances, there have been efforts that tend to utilize multi-source / multi-resolution data and unlabeled data, as well as more noise-tolerant models and learning methods to address these challenges. In general, these efforts can be collectively summarized as below:

(1) *Weakly supervised/Semi-supervised learning to address small, imprecise, and incomplete samples.* Weak supervision assumes the training data to be noisy, imprecise, and unbalanced, while cheap to obtain (e.g., publicly available GIS data). The approaches are often problem-dependent since it aims to build problem-specific heuristics, constraints, error distributions, etc., into the learning models. In RS, this is related to applications dealing with crowdsourcing labels, or labels that have minor temporal differences; semi-supervised learning assumes a small labeled dataset and the existence of a large amount of unlabeled data, the goal of which is to learn latent representations based on the labeled data and uses the limited training data to achieve high classification performance. Special problems or cases of semi-supervised learning include "X"-shot learning (e.g., zero-shot, one-shot, and few-shot), which needs to specify the amount of available labeled data [36,37].

(2) *Transfer learning (TL) and domain adaptation approaches to address domain gaps.* TL is defined in the machine learning domain that assumes knowledge learned from one task could be useful if transferred to another task. For example, a model that learns to perform per-pixel semantic segmentation of scenes can improve human detection. In the RS domain, this largely refers to techniques that minimize gaps in the feature space to achieve a generalizable classifier for data of different sensors or of different geographical locations.

(3) *Use public GIS data or low-resolution images as sources of labeled or partially labeled data.* OpenStreetMap offers almost 80% of GIS data coverage of the globe with varying quality [38], and some local governments release relatively accurate GIS data for public distribution. Researchers had showcased work under this context and achieved conclusions specifically tied to datasets. In addition, as the low resolution labeled data with global coverage are becoming gradually more completed (e.g., National Land Cover Database [1]), these low-resolution labels can be used as a guide to generally address domain gaps of data across different locations for scaling up the landcover classification of VHR data.

(4) *Fusion of multi-modality and multi-view data.* Frequently, there are multiple data sources such as LiDAR (light detection and ranging), SAR (Synthetic Aperture Radar), and nighttime light data. Although these data are mostly unlabeled, they provide additional sources to explore heuristic information and more robust latent representation learning.

## 2. An Overview of Typical Landcover Classification Methods

The RS community has been experiencing a period where DL techniques are more frequently being used, and continue to create "new norms" regarding the workflow to produce various geospatial data products at scale. The conventional dichotomous view of landcover methods that categorizes these methods with either a pixel-based or object-based approach, designated for handcrafted features in traditional (and previously dominant) classifiers such as support vector machines and random forests, may no longer serve these "new norms" and require expansions to accommodate the DL techniques in landcover mapping. The emergence of DL techniques significantly changes the landscape of the technical field for processing high\very-high resolution images. Four primary categories of DL models have been used to extract information from RS images, including (a) scene classification that classifies an image patch and produces a scalar value representing one of the scene types, (b) semantic segmentation that takes image patches as the input and obtains the landcover types for each pixel in the image patch, (c) object detection that processes the image patch to detect object type and produces bounding boxes for each detected object, (d) instance detection and segmentation which not only provides a bounding box as performed by object detection but also delineates the boundary of the object within the bounding box as done by semantic segmentation. Readers are encouraged to refer to [8] for an overview of applications of these four categories of DL models in the RS area. In terms of landcover mapping, scene classification and semantic segmentation models are the most relevant to landcover classification methods. It should be noted that the results of the instance detection and segmentation models may also serve for landcover mapping since their predictions include the semantic segmentation results.

The DL classifiers can differ from traditional classifiers in both input and output formats. The DL classifiers combine the feature learning, feature extraction, and target classification in an end-to-end manner, and take an image patch as input rather than the one-dimensional hand-crafted feature vector required for traditional classifiers such as SVM and RF. Moreover, depending on the type of DL classifiers, the DL classifier can produce a 2D array as an output instead of a scalar value as generated by the traditional classifiers.

In this section, we aim to present a simple way for readers to systematically review the existing landcover methods and recognize their relationships considering the analysis unit. To this end, we created Figure 2 to expand the traditional dichotomous view and to present an overview of the landcover mapping methods inclusive of DL methods. In this figure, the grid icon in the first row represents the RS images in raster format and for simplicity, the band dimension is not shown. The oblique grid denotes the feature extraction area and the purple color represents the pixels to which the classification results are assigned; details are explained in the following subsections.

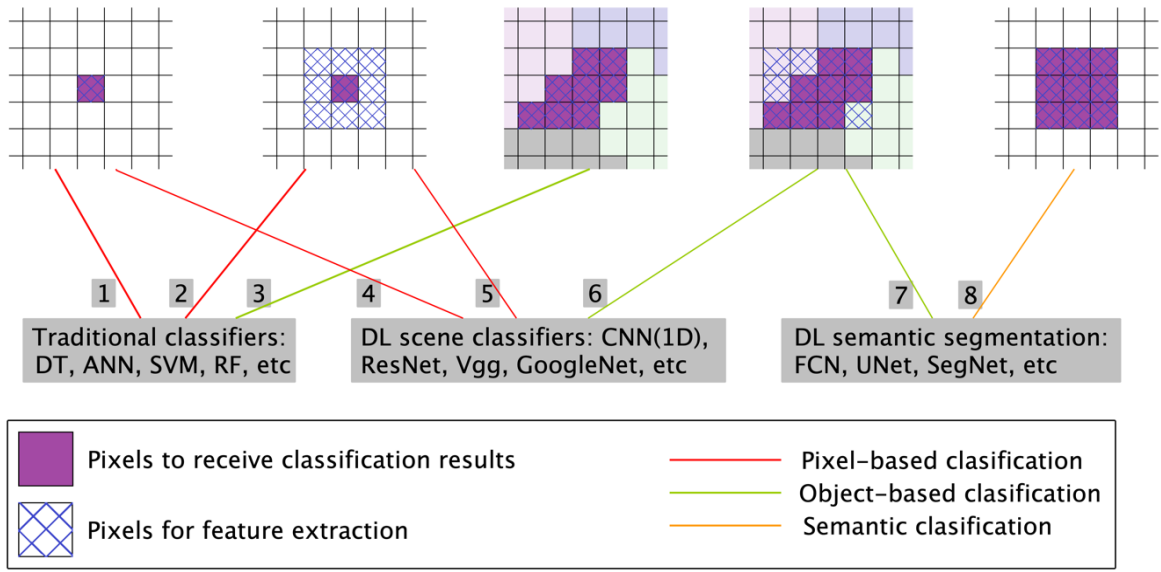

**Figure 2.** An overview of existing land cover classification paradigms in terms of their analysis unit. Purple color represents the classification result assignment area, oblique grid denotes the feature extraction area and different colors in the third and fourth grids represent different objects in the object-based image analysis. Different paths connecting different categories of classifiers and grids are numbered and will be referred to as Path-# when discussed in text.

*2.1. Pixel-Based Mapping Method*

The pixel-based method treats a single pixel as an analysis unit. The simplest implementation of the pixel-based method is shown by Path-#1 in Figure 2, which utilizes the pixel values for all the bands in this pixel location (purple color indicates pixels of concern) to form a feature vector, and employs the traditional classifiers such as random forest (RF) [39,40], support vector machine (SVM) [41,42], artificial neural network (ANN), and maximum likelihood classification (MLC) to perform the classification [43]. In this scenario, one feature vector leads to one scalar value (often an integer), which represents one landcover type as a classification result and is assigned to this pixel location. This procedure is applied to all pixels to produce the landcover map. It should be noted that the bands are not limited to original optical bands, and features such as optical index (e.g., NDVI), and ancillary data (e.g., Digital Height/Surface Model) can also be stacked to the optical bands to derive the feature vectors. To utilize the contextual information for classification, the feature extraction area can be expanded to the neighborhood area of the pixel under consideration as shown by the oblique grid in the raster associated with Path-#2 in Figure 2. In this case, additional features (e.g., standard deviation, gray level co-occurrence matrices [44], etc.) based on the pixels within the neighborhood area can be extracted to represent the texture information of the area surrounding the pixel. The neighborhood area is usually represented by a window centered on the pixel and the window size was shown to be important in impacting the classification results [45]. Readers are encouraged to refer to [46] for more information regarding the pixel-based classification.

DL scene classifiers (e.g., ResNet [47], AlexNet [48], VGG [49], GoogleNet [50], DenseNet [51], SqueezeNet [52], ShuffleNet [53], Inception [54], MobileNet [55]) use image patches as input and output a scalar value as a classification result. The window centered by a pixel can be readily used to obtain an image patch. Therefore, scene classifiers have been commonly used to perform pixel-based classification (Path-#5 in Figure 2). As convolutional neural networks (CNN) rely on the convolutional kernel to slide over the image to perform convolution operations to extract features, the relatively small window (e.g., 3 × 3 or 5 × 5 etc.) used for traditional classifiers to extract features may act as a local feature extractor that is unsuitable for CNN to extract features globally. Hence, the window used for DL classifiers to perform pixel-based classification is comparatively larger, and various sizes of windows were used in existing works, e.g., 5 × 5 [56], 9 × 9 [57], 15 × 15 [58], 400 × 400 [59]. To reduce the computational burden and noise in the resulting classification map, a block of pixels (e.g., 3 × 3, 5 × 5) rather than a single pixel in the window center may be used to assign the classification result [59]. This is a common strategy for landcover mapping tasks that do not require strict delineation of target boundaries (e.g., human settlement mapping). A drawback of this approach is the reduced resolution of the landcover map, since all pixels in the block share the same class label.

Due to the 2D nature of the convolutional kernel in CNN, CNN usually does not apply to single pixels. However, for hyperspectral data with hundreds of bands, the standard 2D convolutional filters can be adapted with a 1D convolutional kernel to take a 1D vector as input (Path-#4 in Figure 2) [60]. In addition, the long feature vector generated from a large number of hyperspectral bands can be folded to form a 2D array, which can be input into standard 2D CNN to perform classification [61]. Readers may refer to [4] for techniques of processing hyperspectral images using DL classifiers.

*2.2. Object-Based Image Analysis (OBIA)*

While the pixel-based method is straightforward to implement and remains popular for medium resolution RS images [41], it introduces undesired effects for high\very-high resolution (VHR) RS images (e.g., salt-and-pepper effects). This is due to the content in a single pixel capturing incomplete information of ground objects. To overcome this issue, object-based image analysis (OBIA) was introduced. The word "object" is somewhat misleading, since it usually does not correspond to one complete object in the real world, e.g., a footprint of a complete building or a tree crown. Instead, one object used in OBIA is simply a group of homogenous pixels, representing the subarea of a landcover class or one real object, thus sometimes these are alternatively termed region-based or segment-based classification. It must be noted that an object is also referred to as a super-pixel among the computer vision community. Objects or super-pixels are generated by image segmentation algorithms. In the RS community, multi-resolution segmentation provided by the eCognition package is usually used for performing OBIA and recently open-source algorithms such as Quickshift [62], and SLIC [63] have been increasingly used. Recently, the Simple Non-Iterative Clustering (SNIC) [64] segmentation algorithm has become extremely popular when OBIA approaches are implemented in the Google Earth Engine. In Figure 2, green paths represent object-based approaches, and different colors in the grid denote different objects.

OBIA assumes that all pixels within an object belong to one single landcover class. This assumption has implications on two aspects regarding the OBIA procedure: the first one is that features are extracted from the object instead of single pixels, since object-based features are assumed to be more informative than the features extracted from single pixels; the second implication is that all pixels within the object will be assigned the same predicted label. Pixel- and object-based methods were compared and systematically analyzed in the RS community, and the consensus reveals that OBIA generates a more visually appealing map with similar or higher accuracy compared with pixel-based methods for high\very-high resolution images [65], while some studies showed that OBIA did not show advantages over pixel-based methods using medium resolution images (e.g., 10 m resolution SPOT-5) [66] in terms of either accuracy or computational efficiency.

Traditional classifiers rely on hand-crafted features extracted from pixels within objects for classification. As compared to the pixel-based method using a neighborhood area to extract features, the object-based method allows the extraction of geometric features (i.e., the features characterizing the shape of objects, such as area, perimeter, eccentricity, etc.), even though those features were less useful in improving the accuracy for supervised classification in some studies [67]. Path-#3 in Figure 2 represents the object-based approach using traditional classifiers, labels are assigned in the area where features are extracted. Readers are encouraged to refer to [68] to review the progress of OBIA.

DL scene classifiers require image patches as input, which can be naturally generated by cropping the image based on the bounding boxes of objects. Therefore, the object-based classification problem is converted to a scene classification problem. This implementation of the object-based method using CNN is shown in Figure 2 through Path-#6, whose associated grid image indicates that the feature extraction and label assignment area are not necessarily equivalent. It should be noted that since the objects are different from each other, the image patches derived from bounding boxes of those objects may have different dimensions. These image patches must be normalized to share the same dimension as specified by the input dimension of DL scene classification models.

The object-based CNN is not limited to the implementation mentioned above and several other types of object-based CNN implementation exist. For example, instead of generating one image patch for an object, users can generate multiple image patches within the object, where the size and quantity of the generated image patches are guided by the shapes and areas of the object, and majority voting was used to summarize all the classification results produced by the image patches within the object to produce the final classification result [58,69-71]. Moreover, the OBIA can be used in a post-processing manner for classification to remove salt-and-pepper artifacts. For example, CNN is firstly implemented within pixel-based approaches (i.e., Path-#5 in Figure 2) to generate a landcover map and then objects are overlaid onto the landcover map, and the label is assigned based on the majority vote, to reduce noise and yield a smoother landcover map [56,72-74]. A similar post-processing strategy via objects can be also applied to maps generated by semantic segmentation models [72,75]. OBIA became popular since its purposeful introduction to address landcover mapping tasks with VHR images[68,75-85], but we observe that the DL semantic segmentation approach, as described below, shows a tendency to replace the OBIA for VHR classification in the future.

*2.3. Semantic Segmentation*

The standard implementation of the semantic segmentation model takes the image patch as input and generates a label for each pixel in the image patch. The procedure corresponds to Path-#8 in Figure 2, and the associated grid image shows that the feature extraction and result assignment are on the same image grid. Semantic segmentation aims to provide a dense and per-pixel label prediction for the image patch; thus, the label assignment or feature extraction is not regarded as dependent on any single pixels or objects, but rather a dense label grid as a whole.

Landcover mapping using semantic segmentation requires a tiling approach; the large-sized RS images are split into overlapping or non-overlapping rectangular image patches. Based on this, labeled image patches are generated separately using a trained model and then stitched back to form a full classification map [86-93]. Given its superior performance in practice, the semantic segmentation models are becoming the most attempted in the RS community using VHR images to generate landcover mapping at scale. For example, building footprints, road network maps, landcover maps, and human settlement maps were successfully generated using the semantic segmentation model. The drawback of the semantic model, as compared to scene-level classification, is that it requires densely labeled training samples, which can be made very expensive. However, utilizing a patch-level label with semantic segmentation models is possible. For example, [94] an assigned background as a class type to all pixels that were outside the object boundary within the image patch, with which a method to train the semantic model using the samples that were collected on object level was presented (Path-#7 in Figure 2) and showed better performance than the scene classifiers. Readers may refer to [95] for more information regarding landcover mapping using semantic segmentation models.

**3. Literature Review of Landcover Classification Methods Addressing the Data Sparsity and Scalability Challenges**

Both the traditional and DL methods for landcover classification require that the annotated training samples are somewhat similar (or geographically close) to the images, which is however difficult to meet, since data of varying geographical regions present vastly different land patterns that are impossible to encapsulate within one single training dataset. To address such challenges, approaches either originated from the machine learning/computer science domain, or the RS domain, were developed to overcome those challenges specific to RS image classification. Here we generally review these efforts under three related/partially overlapped areas: (1) weakly/semi-supervised classification; (2) transfer learning and domain adaptation; (3) multi-source and multi-view image-based classification. Note that we do not intentionally treat traditional and DL classifiers differently, rather we regard them (the classic and deep learning) as models with different complexity. These methods addressing data sparsity and scalability challenges, when applied to RS scenarios, can be briefly described in the following table (Table 1).

**Table 1.** A summary of various learning approaches addressing data sparsity and scalability challenges.

| Methods | Descriptions | Application Scenario in RS Data | Examples of Relevant Works |
|---|---|---|---|
| Weakly supervised/Semi-supervised learning | Semi-supervised learning aims to address tasks where a small set of labeled data and a large amount of unlabeled data are available, while Weak supervision assumes the labeled data to be noisy and contain errors, and the learning methods consider | In RS classification, the noisy inputs are categorized as the following three types: Incomplete: collected samples are too few and biased. Inexact: the form of the training sample does not match with the desired form of classification results. e.g., point samples or scene-level samples vs, per-pixel samples. | [20,36,74,96-117] |
| | the uncertainty level of the available label information. In RS, this is often mixed-used with semi-automation. The readers may refer to the explanations in the texts | Inaccurate samples: error-prone training samples, such as those from crowdsource data (e.g., OpenStreetMap) | |
| Transfer learning and domain adaptation | Transfer learning (TL) is defined as transferring learned knowledge from one task to the other, normally by understanding the distribution of the feature space are different and need to be aligned through domain adaptation methods. | In RS, TL is normally defined as transferring knowledge (e.g., for classification) learned from one dataset and applied to another dataset that is drastically different in geographical location, or captured by different sensors/platforms. This also includes cases where deep models need to be fine-tuned given sparsely labeled data for training. | [20,118-127] |
| Multi-Modal and Multi-view learning | Data fusion methods are general approaches that utilize multiple coherent data sources or labels for performing classification tasks. Multi-view image-based learning is a subset of data fusion approaches that utilize the redundancies of multi-angular images to enhance the learning and is less covered in the literature, which this section will focus on. | Data fusion approaches are widely applicable since multi-modality remotely sensed such as SAR, optical, and LiDAR data, as well as multi-resolution, multi-sensor and multi-view data. The use of multi-view/multi-angular data is very common in photogrammetric collections. Using multi-view images enhances augments information of an area of interest and hence improves the accuracies. | [80,128-131] |

*3.1. Weak Supervision and Semi-Supervision for Noisy and Incomplete Training Sets*

Weakly supervised methods refer to training paradigms that consider training data as noisy inputs. Such "noisy and incomplete inputs" [96] can be categorized into several cases: (1) incomplete: small training sets that are unable to cover the distribution of testing sets; (2) inexact: training sets and labels do not match with the testing sets, e.g., different resolution and details for labels, such as scene-level labels versus pixel-level labels; (3) inaccurate: training sets and labels are not trustworthy and contain errors. In the RS field, weak supervision under the case of "incomplete inputs" is often mixed with semi-supervision, and the only minor differences are that semi-supervision assumes a large amount of unlabeled data to draw marginal data/feature distributions while weak supervision does not require so. Given the large volume of data in RS classification tasks, such minor differences are neglectable, thus we use them interchangeably under this context. These methods primarily focus on training paradigms themselves, such as data augmentation, or formulating regularizations to traditional or DL models, to avoid overfitting. In the following, we review relevant methods in addressing "noisy and incomplete" data on either one or multiple of the above-mentioned challenges.

3.1.1. Incomplete Samples

When samples are incomplete (or not representative) to characterize the data distributions, the learned classifier might be biased even for well-defined classification problems. To differentiate this from domain gap problems, here we limit our review to tasks assuming no large domain gaps, i.e., the training samples ideally capture data points within the distribution of the testing datasets. In classic RS, the representative training samples are usually assumed to be a prerequisite, while it has evolved as an emerging challenge with the data and resolution becoming larger and higher. The line of approaches in addressing this, are to intuitively generate more points through (1) class-specific information, saliency, or expert knowledge, and (2) active learning through statistical or interactive approaches, which are described in the subsections below.

*Generating new samples using saliency and expert knowledge:* A common scheme is to propagate samples alongside the dataset based on neighborhood similarity [97,98], or saliency maps produced with these limited samples, followed by a re-learning scheme [99,100]. The neighborhood similarity-based approach generally assumes that the connected pixels around the sparsely labeled pixels might share the same label, thus can be added to the training sets. This is particularly useful for applications with incomplete samples, for example, road central line data can be obtained through GIS databases, while per-pixel road masks can be too time-consuming to annotate, therefore, additional labels can be added based on saliency maps learned from sparse road pixels from the central lines [104]. A few other studies explore criteria deciding whether these neighboring pixels should be incorporated in the training [97,101-103], and once identified, neighboring pixels can be taken through pixel-centric window [106] or through confidence maps of the post-classification (using the sparse training data). Additionally, RS data comes with its advantage in pattern recognition, in that the multi- or hyper- spectral information provides physical-based cues on different land classes. For example, the use of well-known indices such as NDVI (normalized difference vegetation index) [105], NDWI (normalized water index) [107], BI (building index) [132,133], and shadow index [134], which provides cues of different land classes through spectral or spatial characteristics of the RS data. For example, [20] utilized a series of RS indices as cues to introduce more samples to balance the distribution of samples to yield better classification prediction. Similarly, for wide-area mapping applications, indices are used to indicate coarse and class-specific cues for stratified classification to a finer classification level [108]. Moreover, this type of approach is used in domain-specific classification problems such as those for crop or tree species detection, where vegetation indices are implemented to introduce samples or provide expert knowledge for classification [135,136].

*Generate new samples through active learning:* Active learning seeks to generate data points through pre-classifying the unlabeled data and incorporating data points with confidence, or to allow users to interactively find and add the most important data points that will improve model training. A common paradigm for active learning [137], is to firstly feed the sparse data to classifiers and decide new sample points to be included, based on various criteria involving the use of posterior confidences of the classifier. For example, the unlabeled samples can be firstly classified using available samples, and based on the classification confidence, these unlabeled samples will be ranked based on the top two most probable classes, and manual interactions are needed for ambiguous pixels, i.e., top two classes have similar confidence [138]. Authors of [139] had explored this concept for RS data through both pixel-based or cluster-based strategies, and concluded that active learning can effectively improve classification accuracy. A similar approach based on such a concept can be found in [99], with incremental improvements on feature selections and adopting an iterative scheme to further improve the accuracy for classification tasks where samples are insufficient. It should be noted that since there is a human-in-the-loop component in the process, there are possibilities that new classes can be discovered.

3.1.2. Inexact Samples

Often there exist discrepancies in the form of training samples and the desired representation in the classification. For example, it might be possible that only scene-level (or chip-level) labels are available in certain RS datasets, or in some domain-specific applications such as tree species classification, only individual points marking tree plots (e.g., from Global Positioning System(GPS) points) are available, while the semantic segmentation desires labels for each pixel [106,109], thus under these cases the training samples are inexact. The basic idea for addressing this type of classification problem follows a similar idea by transforming the point-based or scene-based training sample representation to per-pixel and dense probability or saliencies, with the assumption that the scene-level annotations contain most of the contents as described by the label, e.g., a scene with a label "urban scene" is assumed to contain most of the pixels as urban pixels such as buildings or impervious ground. The salient maps are often extracted as one of the feature maps of a deep neural network (DNN), or from classification activation maps (CAM), or posterior classification confidence maps from shallow classifiers. For example, the semantic segmentation approach in [109] used a low-ranked based linear combination of feature maps trained through scene-level data masks, to decide pixel-level content associated with each class; [103] used the CAM layer of a trained network (using sample point centered image patches) to identify locations of red-attacked trees, and they demonstrated that the CAM layer can identify spatial locations at the pixel-level and as a result for deciding pixel-level labels at prediction. A similar concept was tested by [106], in which U-Net [110] was trained by using patches centered around sample points, with the last layers straightened through a global average pooling to match the single-point training data. In a prediction phase, the per-pixel prediction is performed by thresholding the CAM of the last layer (the same size as the original image). In addition to the inexact representation of training and testing data, a common cause of inexact data can be the mismatch of training labels to the desired labels.

As shown in Figure 1, training and testing labels may be at a different level of detail, or the testing sample might have new classes, which can be potentially addressed by zero-shot learning (assuming no label for the new class) [36] or few-shot learning (assuming only a very limited number of sample for the new class) [111,112]. These are relatively less investigated in RS: an example of the former (zero-shot learning) [36] performed a label-level propagation-based language processing models to derive new labels of unseen classes, and an example of the latter adopted the metric-learning strategies by learning a distance function using a large image database [140,141], and by implementing the distance function, it could thus perform predictions. Both of these examples have demonstrated certain levels of improvements [111,142]. Few-shot learning has been recently investigated in RS community for scene classification and showed promising results [143,144], but its application for semantic segmentation model for RS images remains relatively underexplored.

3.1.3. Inaccurate Samples

In an RS context, inaccurate samples mainly refer to cases where the samples contain a certain level of errors or inconsistencies, such as those aggregated from open or crowdsourcing data (e.g., OpenStreetMap [25]), or data labeled from low-resolution data [96]. OpenStreetMap (OSM) is one of the most investigated data sources for classification studies, as it was reported to cover approximately 83% of road networks (as of 2016) [145] and 40% of a street network [38]. Most of the existing studies assume OSM to be error-prone, thus various strategies were developed to refine the labels. For example, [74] explored a data fusion and human-in-the-loop approach, which fused extracted vector data from OSM and overlays with satellite images for visual checking and sample selection, to perform quality control of the labels before training; [113] automated this procedure by using several common RS indices including NDVI [105], NDWI [107] and MBI [132], to remove inconsistent labels using heuristic rules. Authors of [114] employed a simple procedure to refine per class labels: by assuming registration errors of the OSM and the data, they first eroded the OSM labels through binary morphological operations, and in a second step, they performed a clustering procedure on pixels, and labeled them based on the OSM. Both of these studies [113,114] used shallow classifiers. It should be noted that in their studies, the road vectors were considered to be sufficiently accurate and were directly imposed on the final results. DL-based approaches applied to the OSM data [115,116], often consider that the volume of OSM for training is sufficiently large to avoid overfitting problems in DL, thus there is only very light, or no pre-processing applied to the OSM data before training. Alternatives, in addressing these data inaccuracies, the DL methods build loss functions that inherently consider data inaccuracy, such as data imbalances and label noisiness. For example, the work of [115] used pre-defined empirical weights on the losses of different classes (i.e., smaller weight for background and bigger weights for building and road classes) to address the label sparsity. The work of [116] used a similar strategy in their loss function for its application of map generation using generative models. Both works [115,116] reported satisfactory classification results in their applications. There are also other works that develop ad hoc solutions specific to the type of crowdsourcing data or the OSM data. For example, the work of [145] directly utilized OSM data as tile-level input to train a random forest classifier for predicting region-level socioeconomic factors, which reported satisfactory results without any pre-processing of the OSM; the work of [117] reported an application of using crowdsourcing geo-tagged smartphone data for crop type classification, and they reported that the data were manually cleaned for DL.

*3.2. Transfer Learning and Domain Adaptation for RS Classification*

As mentioned above, transfer learning (TL) in machine learning is broadly defined as transferring learned knowledge from one task to other related tasks, for example, applying information learned from the semantic segmentation task to the human recognition task [146]. In RS classification, this often explicitly refers to training a model (or classifier) using one dataset and applying the adapted model to another dataset (with no or very few labels). This is particularly needed in the context that the two datasets stem from different geographical regions, or are collected from different sensors and platforms [19]. In this case, feature distributions extracted from the dataset are different, and it requires domain adaptation methods to minimize the differences of the features across different datasets. In RS classification, the domain with labeled data is defined as the source domain, and the domain with no or very few labels (as compared to the source domain) is regarded as the target domain. There are two types of TL applications when applied to RS applications: (1) domain adaptation and (2) model fine tuning.

3.2.1. Domain Adaptation

Domain gaps are defined as the differences of feature distributions between source and target datasets. This is often regarded as the major cause of the generalization issues among machine learning algorithms. Figure 3 illustrates a simple example that draws the distribution of radiometric values of different datasets: Figure 3a draws typical marginal distributions of the radiometric values of two datasets, which shows a systematic bias; Figure 3b draws the conditional distributions (per class feature distribution) of a two-dimensional feature, which shows that, for the same class, the feature values appear differently. These differences made it different for a classifier trained from the source data, to achieve satisfactory results in the target data. The domain adaptation (DA) methods aim to minimize such differences between feature distributions, which can be either performed explicitly on the feature level (through both handcrafted or learned features), or implicitly performed in latent spaces during the learning process. In the following subsections, we introduce a few typical DA methods used in RS. (1) Feature-level domain adaptation; (2) Domain adaptation through DL-based latent space; (3) Generative adversarial Network (GAN) [147] based domain adaptation.

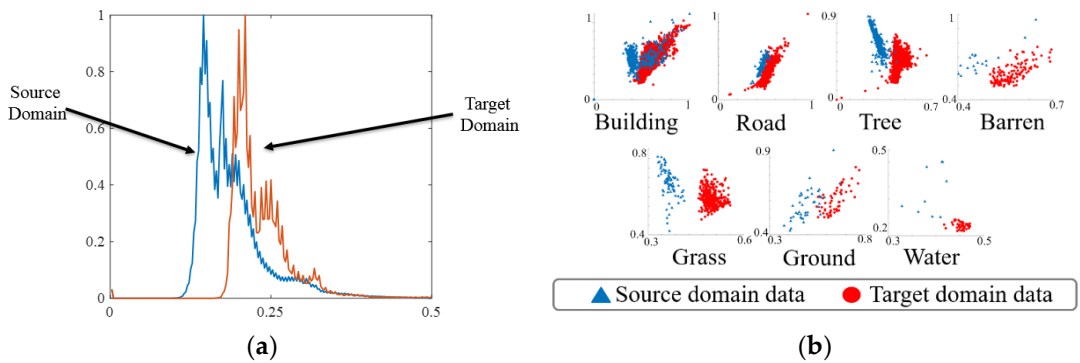

**Figure 3.** An intuitive visualization of typical domain gaps and misalignment of feature distributions between the source and target domain: (**a**) Marginal distribution (feature distribution over all the classes) of radiometric values; (**b**) conditional distribution of a typical two-dimensional feature of the two domains showing the misalignment on different object classes, x and y axis of these subfigures plots are appropriately adjusted for best visualization.

*Feature-level domain adaptation:* The feature-level DA refers to approaches that minimize the feature distributions between the source and target domain. DA approaches may minimize either marginal or condition feature distributions, or both, depending on the availability of these distributions in the target data. Normally aligning the marginal distributions assumes the algorithm is completely agnostic to label space of the target domain, while aligning the conditional distributions requires part of the labels of the target domain data to be known. A DA algorithm proposed by [148] was regarded as one of the simplest methods, which produced copies of features based on limited labeled samples from the target domain, to improve the accuracy for target domain classification, which reported to have achieved competing results in many classification tasks. Since this simple algorithm requires some labeled data in the target domain, [118] proposed another "simple" DA algorithm termed CORAL (CORrelation ALignment) that does not require any labeled target domain data, as it aligns the marginal distributions of the features in the source and target domains through minimizing the differences of the second-order statistics of the distributions.

An important line of work in DA, known as reproducing kernel Hilbert space (RKHS) [119] aims to project both the source and the target feature space to a common metric space. In RKHS, the differences of the feature distributions are minimized under a metric defined as maximum mean distance (MMD), and following the minimized distributions, the classification of the target data can be then performed in the new metric space. An improved version, called transfer component analysis (TCA)[120], incorporated feature transformation (so called transfer component) into the MMD minimization, thus yielding better results. Its semi-supervised version was proposed in [120], which minimizes the conditional distribution by assuming only very limited labels in the target domain. However, it was known that minimizing either marginal or the conditional distribution alone does not encapsulate the entire data distribution. Thus, [121] proposed a joint domain adaptation (JDA) method that jointly optimizes the alignment of the marginal and conditional distribution of features, in which the features were transformed under a principal dimensionality reduction scheme. While requiring a longer computational time, JDA appeared to perform better than TCA.

When target domain labels are not available, an intuitive scheme for generating so-called pseudo labels can be applied: It firstly trains a classifier on the source domain data, then performs classification on the target domain data, and finally takes labels with high confidences as the desired pseudo labels from target domains. However, this will inevitably introduce the sample imbalance problem. Researchers [20] proposed a multi-kernel approach that tends to provide the flexibility to weigh multiple kernels to improve the DA in their experiments, they specifically noted this data proposed to balance the target domain labels for training. Different from most of the aforementioned methods which are mainly practiced in tasks other than RS image classification, the work of [20] was applied to the RS contexts and reported an improvement of 47% on combined multi-spectral and DSM (digital surface model) data. Researchers [122] evaluated the semi-supervised TCA (SSTCA) method on a set of RS images, and reported that the SSTCA method, when combined with a histogram matching between source and target images, achieved approximately 0.15 of improvement in terms of Kappa metric.

*Domain adaptation through DL-based latent space:* Most of the aforementioned or earlier methods are designed for traditional machine learning pipelines. There were a few new developments in recent years focusing on domain adaptation for DL at the latent space or embedding vectors. Since DL methods for classification / semantic segmentation require densely labeled images, it will become more challenging to collect even a small number of samples in the target domain, thus the DA method for DL usually assumes no target labels. The basic idea of this type of approach is to obtain the source domain image, target domain image, and the source domain labels as input for training, and build a loss to infer the embedding vectors. These embedding vectors are often shared between the source and target domain data, to implicitly learn domain adaptations in the embedding space. An example of such a method is the DL version of the CORAL algorithm [123], in which the CORAL loss was built to minimize the second-order statistics of the feature maps, resulting in a shared embedding space for feature extraction and classification. Researchers [124] built the loss function as discriminators of the source and domain data in an embedded space consisting of both learned CNN (convolutional neural networks) and wavelet features, such that the embedding space is less semantic to source and target domain data. Research by [149] attributed the domain gaps to be primarily the scale differences; they built a scale discriminator to serve as a trainable loss to be scale-invariant to objects, which infer the gradients back to the embedding space to improve the DA. A similar but slightly more agnostic approach obtained synthetic data for semantic segmentation in an automated driving scenario, in which they trained a discriminator to identify whether the output semantic maps came from a synthetic or real image. This discriminator essentially serves as a loss to infer the parameters in the shared embedding space, therefore, to improve the adaptation of the semantic segmentation tasks.

*GAN-based domain adaptation*: The GAN-based DA methods aim to use generative models that directly translate target domain images to the styles in the source domains. The ColorMapGAN [125] presented such an approach, which assumes no labels from the target domain and turns target domain images into the images of the original domain through a CNN-based GAN, and showed significant improvement in classification accuracy. Since this type of approach is agnostic to classifiers, it works for both traditional and DL models. The authors expanded this work to a more general framework called DAugNet [34], which incorporated multiple sources and target domains, and can consistently learn from examples. Researchers [126] extended the image translation concepts to include four images as the input for semantic segmentation: the target image, target-styled source image, source image, and reconstructed source image from the target-styled source image, in which the latter three associated with the source had labels to backpropagate gradients for training a shared encoder-decoder. In summary, since this type of approach requires a heavy training process for each pair of source and domain, it may face limitations for real-time applications during test time.

3.2.2. Model Fine-Tuning

Fine-tuning is a specific class of approach popularized along with DL methods, which tends to retrain part of a pre-trained network (e.g., ImageNet [150]) using relatively fewer samples. It has become standard practice to (1) adapt classification tasks with few samples; (2) utilize well-pretrained networks as feature extractors. Authors of [31] performed a deep analysis on the transferability of deep neural networks (representatively forward feed networks in this work), and concluded that initializing the networks with pretrained parameters and leaving any number of layers to be trained (depending on the available samples), turned out to be beneficial to achieve high classification accuracy. As a standard practice, this was applied in many RS classification tasks given the lack of labeled data. For example, in the ISPRS semantic segmentation challenge for RS images [151], the majority of the DL approaches use pre-trained networks (often from ImageNet) as a start, and achieved an accuracy beyond traditional approaches. Researchers [127] fine-tuned networks for satellite image classification, and concluded that fining-tuning half of the layers might achieve the same level of accuracy as fine-tuning the entire network, while it could obtain a higher convergency speed. This might be due to well-trained parameters that might serve as a regularization when training models. Other successful applications include (but are not limited to) the Function Map of the World challenge [152], in which solutions for scene-level RS classifications achieved the best results using fine-tuned networks. Authors of [153] used an ensemble of CNN, the training of which started with pre-trained weights and achieved 88.5% in the ISPRS benchmark. Despite the existing work having largely explored fine-tuning as an option to adapt pre-trained networks to address learning tasks with sparse labels, the detailed fine-tuning strategies, for example, freezing how many and which layers, and what regularization constraints to use, are still following a trial-and-error approach. Yet a thorough, and systematic study on fine-tuning approaches for RS applications remain lacking.

Moreover, fine-tuning with labeled samples can follow a self-supervised representation\metric learning procedure, where positive and negative pairs are generated automatically from a large amount of unlabeled images and are used to train a deep learning backbone (e.g., ResNet) to minimize the feature distance of positive pair samples and maximize the feature distance for negative samples [154,155]. Researchers [156] proposed a metric learning model named discriminative CNN(D-CNN) to perform metric learning and obtained state-of-the-art performance on three public RS datasets for scene classification tasks, showing promising results of using self-supervised learning approaches for landcover mapping task. However, it remains to be seen whether self-supervised could also help semantic segmentation using RS images, although this has been confirmed to be beneficial in computer vision community [157].

*3.3. Multi-Sensor, Multi-Temporal and Multi-View Fusion*

The earth observation (EO) data have been growing rapidly owing to the increased number and types of earth observation sensors launched in the past years. The large volume of EO data is heterogeneous in terms of temporal resolution, spatial, spectral resolution, collection angle, or sensor types. In addition to EO sensors, geospatial data is being generated from other sources such as social media and mobile phone locations. Investigating methods to fuse the geospatial data collected across diverse sources or different temporal steps, has attracted increasing attention from the RS community, owing to their complementary characteristics or synergistic effects which can potentially improve the accuracy, temporal resolution, and spatial resolution for the land cover classification results. While a strict definition of taxonomy regarding the data fusion approaches cannot be found in existing literature, data fusion for classification is generally considered at the pixel level, feature level, or decision level, as shown in Figure 4. In Figure 4, two sources are used for simplicity to represent multi-modal geospatial data (e.g., sentinel 1 vs. sentinel 2, spectral vs. LiDAR, EO data vs. social media data, collection angle 1 vs. collection angle 2, collection time step 1 vs. 2, etc.), and it should be noted that the fusion approaches may be extended to multiple sources.

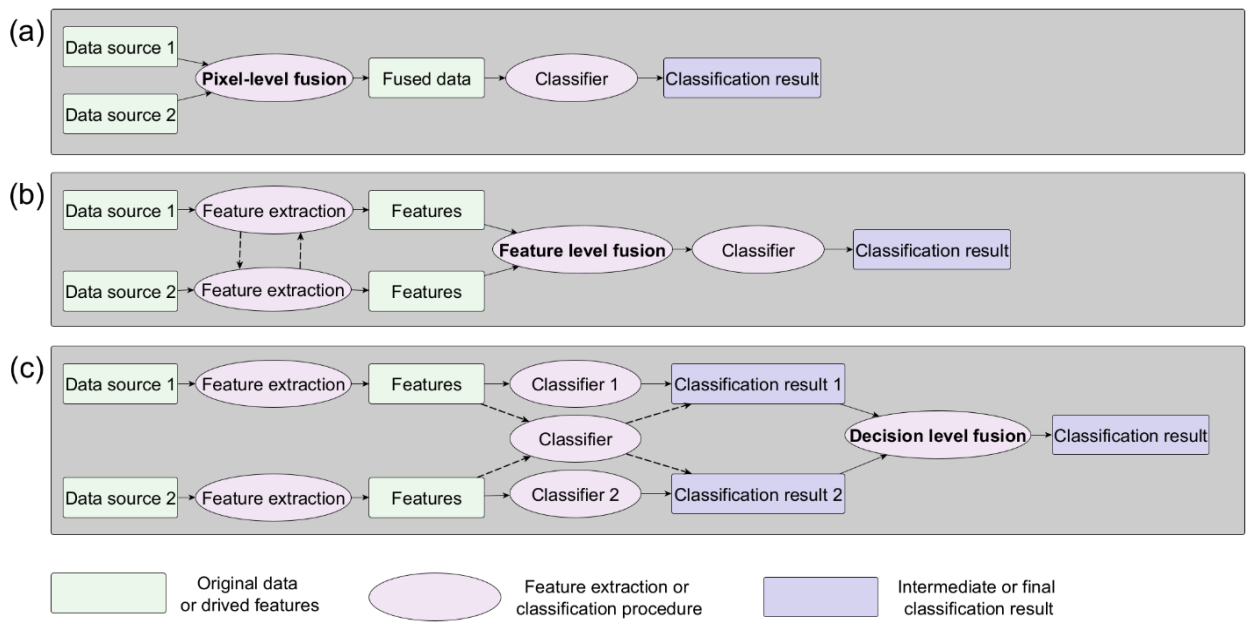

**Figure 4.** Data fusion for classification can be performed at the (**a**) pixel level, (**b**) feature level, and (**c**) decision level. Dashed lines represent alternative paths in contrast to solid lines.

3.3.1. Pixel-Level Fusion

Pixel-level fusion refers to the fusion of information from two sources to generate a new set of data and each band of the new set contains the information from both sources. One pixel-level fusion approach is the well-known principal component analysis (PCA): [158] performed the principal component analysis (PCA) for the Sentinel optical images, replaced with last two PCA components with Sentinel SAR images, transformed the new set of PCA components back to the original format of the optical images as the result of fused data, which was then input into the UNet [110] for crop-type classification. Another category of pixel-level data fusion method relies on pan-sharpening techniques specifically designed to fuse the high spatial resolution, single-band panchromatic with low-spatial resolution multi-band images to generate high-spatial resolution multi-band images. Authors of [159] conducted pan-sharpening algorithms to generate high resolution multiple spectral images, which were used for ice-wedge polygon detection with Mask R-CNN [160]. Eight pan-sharpening algorithms were compared in this study and it was concluded that the performance of the pan-sharpening algorithm was scene dependent, and indicated that the fusion algorithms that preserve spatial characteristics of the original PAN imagery benefit the DL model performances. Researchers [161] proposed a hybrid fusion approach that not only performs pan-sharpening to generate high spatial multi-band images, but also proposed a module named correlation-based attention feature fusion (CAFF) to fuse the features from both sources for image classification. Pan-sharpening itself is an active research topic, for which there are a few works [162-164] providing comprehensive reviews. In addition to PCA and pan-sharpening techniques, other approaches were developed. For example, [165] proposed a new approach that constructed K filters corresponding K data sources to convert the K sources of data to a fused dataset with K bands called latent space, which was then used to map local climate zones with SVM classifier.

In addition to pixel-level fusion that operates in the spatial–spectral domain, it can also be conducted in the spatial–temporal domain, which primarily aims to increase the temporal resolution of high spatial resolution images [35,166]. For example, [166] developed an approach based on deep learning techniques to generate temporally dense Sentinel-2 data at 10-m resolution with information of Landsat-2 images. Temporal–spectral fusion can be decoupled with the landcover classification task and is beyond the scope of our study for a detailed review. Readers can refer to a review article on this topic [167] for more information.

3.3.2. Feature-Level Fusion

Feature-level fusion is the most common approach to fuse multi-modal data for landcover classification due to its simplicity in terms of implementation and concept. Among other types of feature fusion approaches, concatenation of handcrafted features is the most common in the existing literature. For example, features derived from LiDAR and hyperspectral images were concatenated to map land covers with random forest classifier, as is shown in the work of [168], where such approaches were applied to the Houston, Trento, and Missouri University, and University of Florida (MUUFL) Gulport datasets. Other types of features such as spectral indices, texture metrics, and seasonal metrics, can be extracted and concatenated with location-based features to map the human settlement using a random forest classifier [169].

In recent years, the number of papers that utilize the DL network to fuse the features has surged. Given datasets of two sources, a common DL architecture for fusing the features extracted from the two sources has two streams, with each stream taking one type of data as input and the features extracted from two streams in the last layer, followed by concatenation and being fed into a fully connected layer for classification. Ref. [170] adopted this DL architecture to fuse LiDAR and spectral images for landcover classification in an urban environment. To increase the feature representation, more than one type of DL network can be used to extract features from one data source. For example, long short-term memory (LSTM) and CNN were both used to extract features from social sensing signature data, which were concatenated with features extracted with CNN from spectral images to classify urban region functions [171]. The feature fusion may also be performed in intermediate layers in addition to the last layer [172-175], which is indicated by the dashed arrow lines in Figure 4b. In addition to concatenation, features can also be fused by maximum extraction operation, i.e., for each position in the feature vector, selecting the maximum values among the features extracted across all the data sources [175].

Natural Language Processing (NLP) algorithms aim to extract meaningful information from a sequence of words. One sentence consists of multiple words, and those words can be represented with a 2D array, where the length of the rows is the same as the number of words of the sentence and each row corresponds to the embedding associated with one word. Moreover, the multiple-temporal multi-spectral RS images can be represented with a 2D array as represented for one sentence with each row corresponding to the spectra values at a one-time step. Due to this connection of RS time series with NLP problems, many NLP algorithms have been adapted as a feature fusion approach to process the RS time-series data to generate the crop map. For example, the commonly used NLP algorithms Recurrent Neural Network(RNN) and Long Short-Term Memory (LSTM) [176,177] have been employed to process the 12 monthly composites of 9 bands of sentinel-2 time-series data for crop mapping task [178]. The latest state-of-the-art NLP algorithm named transformer [179] was also employed to process the time-series RS data for crop mapping [178,180]. For example, [181] proposed a model that combines LSTM with a transformer attention module to process 23-time steps of a 7-day composite of 6 spectral bands of Landsat Analysis Ready Data (ARD) to generate crop data layer (CDL). The CNN module has a good capability of extracting features from the scene context, but none of those multi-temporal approaches have attached a CNN module to the sequence processing module for crop mapping.

3.3.3. Decision-Level Fusion

Decision-level fusion is performed after each classifier leads to a decision from one data source and the decision can be represented as a class label or class probability (Figure 4c). The majority voting [182,183] or the summation of class probability across different data sources [175] can be conducted to implement the decision fusion. Additionally, a final decision can be obtained based on the confidence level of the predicted class label [184]. Figure 4c shows two approaches for training the classifiers: one (solid line arrows) is to train separate classifiers for different data sources and the other (dashed line arrows) is to train one classifier using all the available sources. It should be noted the decision-level fusion used in [175] is different from the one indicated in Figure 4c since the decision vector used in [175] is more similar to the features used in feature fusion by DL networks.

3.3.4. Multi-View Fusion

The multi-view based classification method we introduce here refers to photogrammetric images or multi-view/multi-date images covering the same region. By building the relationship between the multi-view images and their respectively generated digital surface models (DSM), users can build pixel-level correspondences to share labels among these images, and at the same time utilize the spectral redundancies to improve classification results [81]. Multi-view data were considered detrimental to the classification accuracy, as multi-view images needs to be normalized to the nadir view to improve the classification accuracy [185]. However, recent studies show multi-view data contain extra information compared with single-view RS images and the redundant information contained in multi-view images can be used to improve the classification.

Early efforts employ multi-view reflectance to determine the parameters of the Bidirectional Reflectance Distribution Function (BRDF), which characterizes multi-view reflectance patterns. Then, the BRDF parameters are used as a proxy of multi-view information and are concatenated with other features or used alone as the feature vector for traditional landcover classifiers [81,186-191]. In addition to BRDF, other multi-view feature extraction methods were developed for the traditional classifiers to improve landcover mapping accuracy. For example, [192] extracted angular difference among multi-view images for urban scene classification, [193] applied bag-of-visual-words(BOVW) to multi-view images for urban functional zone classification, and [194] proposed Ratio Multi-angular Built-up Index (RMABI) and Normalized Difference Multi-angular Built-up Index (NDMABI) for built-up area extraction.

BRDF-based methods is computationally expensive and other methods mentioned above are tailored for specific applications with specific multi-view datasets. Researchers [183] proposed a simpler yet more efficient and general approach, which fuses multi-view objects by training the classifier using multi-view objects and obtains the final classification result by the majority-voting of inference results from multi-view objects, which works similarly as indicated in Figure 4c. This method accommodates the varied number of views and shows substantially higher efficiency for fusing multi-view information compared to the BRDF model. In addition to traditional classifiers, the method is applicable to DL scene classifiers. It was demonstrated that the convolutional neural network benefits more from the redundancies of multi-view data than traditional classifiers (e.g., SVM, RF) for improving the accuracy of landcover mapping with this method [80]. Finally, multi-view information can also be used with the semantic segmentation model. For example, [115] adapted the semantic segmentation model to allow it to obtain the stacked multi-view tensors in the model for semantic mapping using multi-view information and demonstrated that their method gave better performance than methods that used different views separately.

## 4. Final Remarks and Future Needs

The efforts in different classes of learning methods for image classification generally originate from the machine learning / computer science domain and present a varying degree of preferences in the RS domain in recent literature, while existing attempts in addressing different aspects of the RS landcover classification problems for very high resolution (VHR) remain limited and novel applications, adaptation, and reformulation of these contexts into RS problems are still greatly needed. We consider the landcover classification as a highly disparate problem, and on one hand, the solutions and models may need to accommodate available data and scene content, and on the other, new data sources and novel use of existing and open data may be explored for incorporation to improve landcover classification of VHR images. Considering the multi-complex nature of this problem, this paper, in contrast to other existing feature/classifier specific reviews, has provided a comprehensive review on recent advances in landcover classification practices, specifically focusing on approaches and applications that directly or indirectly address the scalability and generalization challenges: we first presented a general introduction to commonly used classifiers with an expanded definition of analysis unit to incorporate deep learning paradigms, and then described existing solutions and categorical methods in addressing the need for classification in weak/semi supervision context where samples can be incomplete, inexact and inaccurate, in addition to domain adaptation approaches in the case of model transfer and model reuse in different contexts; finally, we surveyed existing paradigms that explored the use of other types of data for fusion.

Many of the existing works using DL semantic segmentation for landcover classification, have overwhelmingly demonstrated the level of improvement. DL-based semantic segmentation methods have received increasing attention for landcover mapping, and it is likely to replace the OBIA in future, which is the current solution aiming to overcome issues brought about by VHR images but has often been criticized for imperfect results and extra computational costs of the image segmentation procedure. It is expected to continue in the future and ultimately has the potential to become the standard landcover mapping approaches. There exist benchmark datasets for methodology comparison, however, using these methods in practice are far more complicated than standard tests, due to the massive influences of the availability of quality training data, as well as the multi-complex nature of data resulting from multi-sensors and multi-modal outputs.

During the survey, we additionally discovered the consensus that using data fusion for RS classification provides a promising direction and is highly needed; with the number of relevant works continuously increasing, however, since the data sources show disparity across different works, a taxonomy and a comprehensive comparison among those methods remains currently lacking in the literature. Moreover, we noticed that researchers in RS communities do not release codes as often as those in the computer vision\science community, making the benchmarking for data fusion development in RS community more challenging. In addition, current applications of domain adaption, self-supervised training and meta-learning in the RS community have not caught up with the latest technical progress observed in the computer science community, where it is claimed that "the gap between unsupervised and supervised representation learning has been largely closed in many vision tasks" [195]. Considering unlabeled RS images are being continuously collected each day from various sensors covering different temporal and spatial domains, investigating how those unlabeled images can be utilized with self-supervised or domain adaption techniques to train a better supervised model could provide a promising direction.

It is generally agreed that solutions for landcover classification problems are still ad hoc in practice, requiring training data with good quality. However, we observe a trend where more research works are shifting their gears from achieving high accuracy with more complex DL models, to achieving more general and global-level classification. This is fueled by the need to, (1) address the challenge of data sparsity, inaccuracy and incompleteness; (2) harness the ever-growing number of sensors with different modality to achieve solutions free of moderation by experts.

To this end, based on this review, we provide a few recommendations for the future works in this research line: (1) Developing domain adaptation approaches taking advantage of the unique characteristics of RS data, such as their diversity in land patterns of different geographical regions, the availability of low-resolution labels for semi-supervised DA, available height information globally, as well as physics-based spectrum signatures in the RS world. (2) Exploring the underlying mechanisms of spectrum diversity across different sensors, to achieve inter-sensor calibration prior to classification. (3) Alleviating the cost of training sample collection for semantic segmentation using non-standard, crowd-sourced means and developing methodologies that standardize use of common crowd-sourcing data (such as OSM) for classification, and easier means to access globally available labels for training. (4) Evaluating the extra benefits brought about towards transfer learning by big databases of labeled RS images compared with computer vision datasets. (5) Establishing more comprehensive benchmark datasets for assessing the generalization capabilities (e.g., few-shot learning task, domain adaption task) of DL solutions especially for the semantic segmentation models; (6) Analyzing the roles of self-supervised learning, active learning and meta-learning in reducing the cost of using deep learning semantic segmentation models for landcover mapping.

**Author Contributions:** Qin, R., and Liu, T., both made significant contributions to this paper.

**Funding:** No funding was used to support the work for this project.

**Institutional Review Board Statement:** Not applicable.

**Informed Consent Statement:** Not applicable.

**Data Availability Statement:** Not applicable.

**Conflicts of Interest:** The authors declare no conflict of interest.